\documentclass{article}
\usepackage{spconf,amsmath,graphicx}
\usepackage{cite,comment}
\usepackage{xcolor}

\title{Wav2vec-based Detection and Severity Level Classification of Dysarthria from Speech}
\name{Farhad Javanmardi, Saska Tirronen, Manila Kodali, Sudarsana Reddy Kadiri, and Paavo Alku}
\address{Department of Signal Processing and Acoustics, Aalto University, Finland.}
\begin{document}
\ninept
\maketitle
\begin{abstract} 

Automatic detection and severity level classification of dysarthria directly from acoustic speech signals can be used as a tool in medical diagnosis. In this work, the pre-trained wav2vec 2.0 model is studied as a feature extractor to build detection and severity level classification systems for dysarthric speech. The experiments were carried out with the popularly used UA-speech database. In the detection experiments, the results revealed that the best performance was obtained using the embeddings from the first layer of the wav2vec model that yielded an absolute improvement of 1.23\% in accuracy compared to the best performing baseline feature (spectrogram). In the studied severity level classification task, the results revealed that the embeddings from the final layer gave an absolute improvement of 10.62\% in accuracy compared to the best baseline features (mel-frequency cepstral coefficients).

\end{abstract}
\begin{keywords}
Dysarthria, Severity level classification, Wav2vec 2.0, MFCCs.
\end{keywords}
\vspace{-0.3cm}
\section{Introduction}
\label{sec:intro}
\vspace{-0.1cm}
Dysarthria is a neuro-motor disorder caused by neurological damage of the motor component of speech production. Dysarthria occurs either due to a neurological injury (i.e., cerebral palsy, stroke) or due to a neurodegenerative disease (i.e., Parkinsons’s disease, Huntington’s disease). Dysarthric speech is often associated with atypical speech prosody, and reduced tongue flexibility and imprecise articulation, all of which impact speech intelligibility \cite{duffy2019motor}.  Assessment of speech intelligibility is essential in distinguishing the progression of dysarthria. 

Speech assessment is  generally performed in voice clinics by speech-language pathologists who conduct intelligibility tests to identify the potential presence of dysarthria as well as its severity level \cite{kent1989toward}. Subjective intelligibility tests are costly and laborious, and they are prone to biases of pathologists due to their familiarity and experience with patients. Hence, the design of an objective method for the severity assessment of dysarthric speech is important. The assessment of dysarthric speech is carried out in two phases consisting of (1) the identification of the presence of dysarthria and (2) the estimation of the severity level of the disease.  Both of these phases are important diagnostic steps that are needed to make clinical decisions on medication and therapy of the patient. This work focuses on both of the above-mentioned phases by studying speech-based detection and severity level classification of dysarthria. In both of these topics, the current study focuses on the use of the pre-trained wav2vec model to extract the features \cite{NEURIPS2020wav2vec2}.

Automatic detection and severity level classification of dysarthria from speech is enabled using data-driven approaches based on supervised learning. This involves building machine learning models that are trained using speech data which has been collected from patients and which has been labelled by speech-language pathologists. Many automatic dysarthria detection and severity level classification methods are based on acoustic features that characterize the salient aspects in the production of dysarthric speech \cite{kim2008dysarthric,kim2015automatic}. Abnormal variations of pronunciation, prosody and voice quality were investigated by using the sentence-level features in \cite{kim2015automatic}. In \cite{gurugubelli2019perceptually,gurugubelli2020analytic}, authors investigated the single frequency filtering -based features for dysarthric speech detection and 4-class intelligibility classification (very low, low, medium, and high). In \cite{kadi2016fully}, auditory distinctive features (based on models of mid-external ear and basilar membrane) were proposed for the assessment of dysarthria. Their study also showed that the combination of auditory features with mel-frequency cepstral coefficients (MFCCs) improves the intelligibility estimation of dysarthric speech. The short and long-term temporal measures based on the log-energy dynamics and auditory inspired modulation spectral features were investigated for dysarthric speech intelligibility assessment in \cite{falk2012characterization}.  Glottal source features along with the OpenSmile features \cite{opensmile} were investigated both in detection of dysarthric speech as well as in classification of the intelligibility of dysarthric speech in \cite{narendra2018dysarthric,narendra2019dysarthric}. Linear weighted combination of articulation, phonation, and prosody features of speech were used in intelligibilty estimation in \cite{rong2016predicting,de2002intelligibility}. Recently, in \cite{chandrashekar2019spectro} authors explored the use of different spectro-temporal representations, such as the spectrogram and mel-spectrogram, and convolutional neural networks in intelligibility assessment of dysarthric speech. 

In a few recent years, pre-trained neural network models have become popular for various speech technology tasks, such as automatic speech recognition (ASR), speaker recognition and emotion recognition ~\cite{NEURIPS2020wav2vec2,hernandez2022cross,mohamed2021arabic,vaessen2022fine}. In the current study, we take advantage of an effective pre-trained model, wav2vec, in speech-based detection and severity level classification of dysarthria. The topic is motivated  by recent findings reported in recognition of pathological speech that have shown good performance for the wav2vec models ~\cite{hernandez2022cross,grosz2022wav2vec2}.  
These models are generally first pre-trained in an unsupervised manner on large speech datasets and then fine-tuned on a small set to perform the required task. In this work, we experiment with the wav2vec model shared on HuggingFace \cite{wolf2019huggingface}. 

The main contributions of the study are:
\begin{itemize}
    \item Layer-by-layer analysis of the utility of the wav2vec 
    features in the detection of dysarthria (healthy $vs$. dysarthric) and in the classification of the severity level (very low, low, medium, and high) of dysarthria from speech.
    \item Systematic comparison between 
    popularly used spectral features and the wav2vec features.
\end{itemize}

\vspace{-0.3cm}

\section{The detection and severity level classification systems}
\label{sec:classification_setup}
\begin{figure}[h]
\centering
\includegraphics[width=\columnwidth,height=4cm]{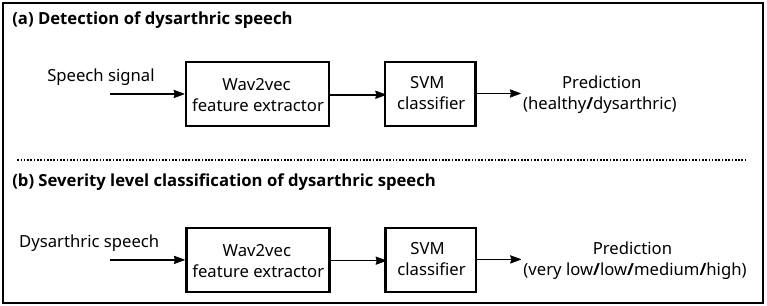}
\caption{\label{fig:fig1} The proposed systems for (a) detection of dysarthric speech and for (b) severity level classification of dysarthric speech.}
\end{figure}

\vspace{-0.15cm}
This work studied the following two classification problems: (1) a binary classification problem to distinguish dysarthric speech from healthy speech (i.e., detection problem), and (2) a multi-class classification problem to classify the severity level of dysarthria from speech to 4 classes (very low, low, medium, and high). In both problems, a pre-trained wave2vec 2.0 \cite{NEURIPS2020wav2vec2,hernandez2022cross} model is used as a feature extractor, and a support vector machine (SVM) is used as a classifier. A schematic block diagrams of the systems built for the two aforementioned problems is shown in Figure~\ref{fig:fig1}. The following two sub-sections describe the feature extraction and classification methods in more detail.

\subsection{Pre-trained feature extractor}
\label{ssec:features}
In building the classification systems for both problems, we use the pre-trained wav2vec 2.0 \cite{NEURIPS2020wav2vec2} model as a feature extractor. 
We utilize Facebook's wav2vec 2.0 model that has been trained with 960 hours of audio from the Librispeech corpus \cite{NEURIPS2020wav2vec2}. The model was originally trained to be used in ASR, 
which implies that the final layers of the network have learnt to extract embeddings that contain information about phoneme identity of speech. However, the embeddings from the first layers of the network also contain information about phones, which makes them useful features also for many other speech-related tasks than ASR \cite{gauder2021alzheimer}.

We take use of the inputs of the first transformer block of the context network of the wav2vec model, as well as of the output embeddings of each of the 12 transformer blocks. As the embeddings are extracted for each non-overlapping 20 ms frame of the signal, we compute the average over the frames to obtain the final feature vectors. The dimensionality of the embeddings is 768, which is also the dimensionality of the features that we finally use in our classifiers. These features will be referred to as the wav2vec features in this paper. To refer to the individual layers, we use the layer numbers, i.e., "wav2vec-N" refers to the features associated to the $N$th layer of the model.

\subsection{Classifiers}
\label{ssec:classifiers}
In this study, a binary SVM classifier is used to distinguish between healthy and dysarthric samples. 
For the multi-class classification between the four severity levels (very low, low, medium, and high), SVM with the one-vs-one architecture \cite{scikit-learn} was used. Both in the binary classification and in the 4-class classification, we used radial basis function as kernel, and a regularization parameter value of 1. For gamma, we used scaling according to the dimensionality and variance of data, written as \( \gamma = 1 / (D \cdot Var(X)) \), where $Var(X)$ is the variance of the training data, and D is the dimension of the feature vectors.

\section{Experimental setup}
\label{sec:experiments}

\subsection{Database of dysarthric speech}
\label{ssec:databases}
This study uses the UA-speech database of dysarthric speech \cite{kim2008dysarthric}. The database consists of 765 isolated words recorded in three blocks (B1, B2, and B3) by 15 dysarthric speakers (four female and eleven male) diagnosed with cerebral palsy and 13 healthy controls (four female and nine male). The overall intelligibility of each of the dysarthric speakers has been assessed using the subjective evaluations with 5 native listeners. Based on the intelligibility ratings, the dysarthric speakers have been grouped into four severity categories: very low (4 speakers), low (3 speakers), medium (3 speakers) and high (5 speakers). Each block contains 255 words in which 155 words are common to all three blocks, and the remaining 100 words differ across the blocks. An eight-microphone array was used for speech recording. Speech was sampled at 16 kHz and each microphone was spaced at intervals of 1.5 in. The current study was carried out using the speech utterances from all three blocks of each speaker, recorded by microphone number 6 of the array. More details of the UA-speech database can be found in \cite{kim2008dysarthric}.

\subsection{Baseline features used for comparison}
In order to compare the performance of the wav2vec features, three popularly used spectral features (spectrogram, mel-spectrogram and MFCCs) were considered as they were shown in \cite{gurugubelli2020analytic} to provide good discrimination between dysarthric and healthy speech. All these feature representations are derived using the Librosa toolkit \cite{mcfee2015librosa}.

\subsubsection{Spectrogram}
The spectrogram features were computed by taking the logarithm of the amplitude spectrum that was estimated using a 1024-point fast Fourier transform (FFT) with the Hamming windowing in frames of 25 ms with a shift of 5 ms. Finally, the features of the spectrogram were averaged over the time axis yielding a 513-dimensional feature vector per utterance.

\subsubsection{Mel-Spectrogram}
The mel-spectrogram features were computed by applying a mel-filterbank with 80 filters on amplitude spectrum. The resulting mel-spectrogram was mapped to a decibel scale through logarithm. By averaging the mel-spectrogram features the over time axis, a 80-dimensional feature vector was obtained per utterance. 

\subsubsection{MFCCs}
To compute the MFCCs, the discrete cosine transform (DCT) was used to transform the mel-scale spectrum into mel-cepstrum. The first 13 cepstral coefficients (including the $0^{th}$ coefficient) were considered and their delta \& double-delta coefficients were computed, which yielded a 39-dimensional MFCC feature vector per utterance.

\begin{figure*}[h]
\centering
\includegraphics[width=1.1\textwidth,height=7cm,trim={2cm, 0cm, 0cm, 1cm,clip}]{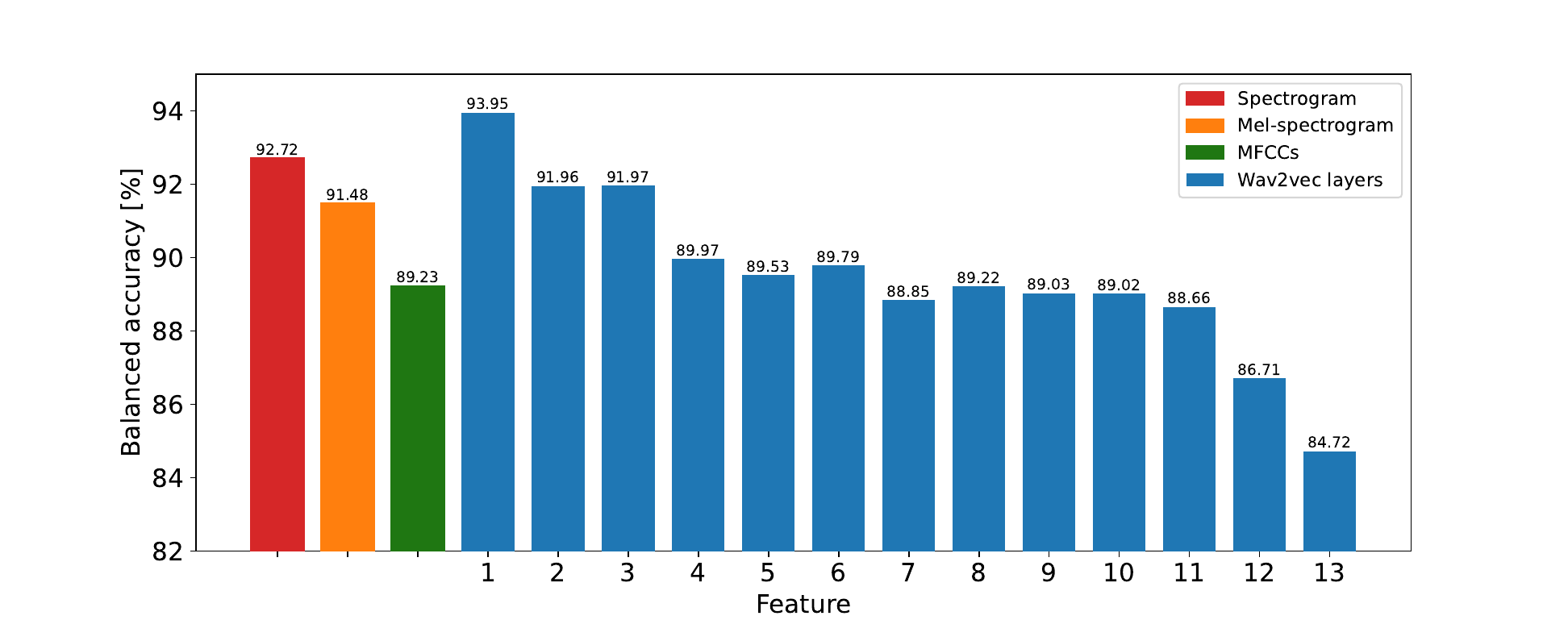}
\vspace{-0.8cm}
\caption{\label{fig:fig2} Detection accuracy of dysarthric speech for the three baseline features (spectrogram, mel-spectrogram and MFCCs) and for all the 13 wav2vec features. The blue bars represent the features derived from the wav2vec model, with the tick labels indicating the index of the corresponding layer. Heights of the bars represent the mean accuracy.}
\end{figure*}

\begin{figure*}[h]
\centering
\includegraphics[width=1.1\textwidth,height=7cm,trim={2cm, 0cm, 0cm, 1cm,clip}]{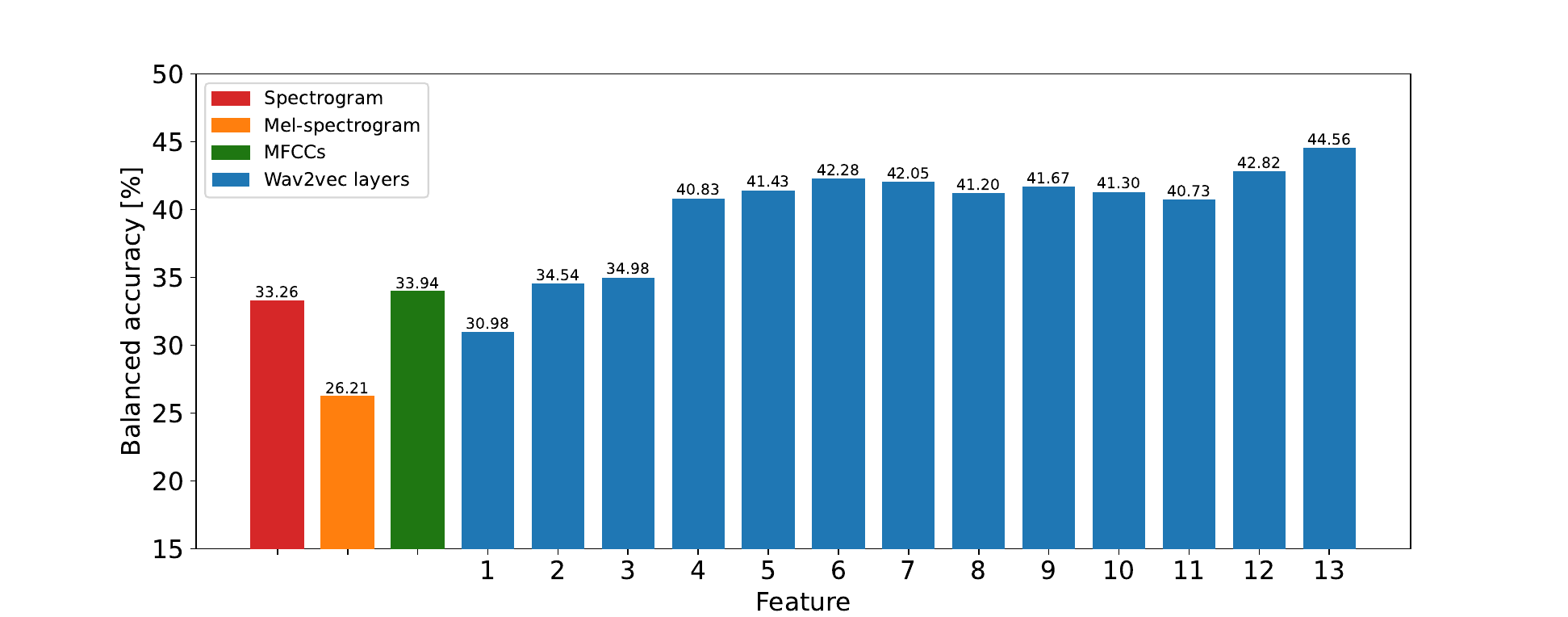}
\vspace{-0.8cm}
\caption{\label{fig:fig3} Severity level classification accuracy of dysarthric speech for the three baseline features (spectrogram, mel-spectrogram and MFCCs) and for all the 13 wav2vec features. The blue bars represent the features derived from the wav2vec model, with the tick labels indicating the index of the corresponding layer. Heights of the bars represent the mean accuracy.}
\end{figure*}

\subsection{Training and testing}
\label{ssec:evaluation}
The binary detection experiments were conducted with leave-one-speaker-out (LOSO) cross-validation, where speech signals of one speaker was considered as test data and speech signals of the remaining 27 speakers were used for training the SVM classifier. Both training and testing data were z-score normalized using the mean and standard deviation of the training data. In each iteration, the evaluation metrics were saved. This process was repeated for 28 iterations (equaling to the number of speakers), and finally the evaluation metrics were averaged over the 28 iterations. 

For the severity level classification experiments, three dysarthric speakers were left out (one male speaker from "very low" level of intelligibility and two male speakers from "high" level of intelligibility) to have the same number of dysarthric speakers in each class. Experiments were carried out using the 12 remaining dysarthric speakers. In each iteration, speech signals of four speakers (one from each class) were considered as test data and speech signals of the remaining speakers were used to train the classifier. By considering one speaker from each class for testing the SVM classifier, a total of 81 (3*3*3*3) iterations (training and testing process) were performed.

\subsection{Evaluation metrics}
In order to evaluate the performance of the dysarthria detection systems, the following four commonly used evaluation metrics were considered in this study: accuracy (ACC), sensitivity (SE), specificity (SP), F1-score (F1) apart from the confusion matrices. For the severity level classification systems, mean accuracy and class-wise accuracies were used for assessing the performance of the systems.

\section{Results}
\label{sec:results}
This section reports the results obtained using the wav2vec features and the baseline features. The results are reported in Section \ref{sec:res_binary} for the detection problem (i.e., the binary classification problem) , and in Section \ref{sec:res_multi-class} for the 4-class severity level classification problem. 

\subsection{Detection of dysarthric speech}
\label{sec:res_binary}
The obtained accuracies of the binary detection experiments are shown in Figure \ref{fig:fig2} for the three baseline features and for all the 13 wav2vec features. Table~\ref{tab:binary_res} presents the results for the baseline features and for the three best wav2vec features by showing all the four selected metrics. It can be observed that the wav2vec-1 outperformed all the other features in all metrics. In particular, when compared to the best performing baseline feature (spectrogram), wav2vec-1 showed an absolute improvement of 1.23\% in accuracy. This finding indicates that the first embedding layer of the wav2vec model learns generic speech representations that are useful for distinguishing dysarthric speech from healthy speech.

Confusion matrices for the dysarthria detection experiments are displayed in Table~\ref{tab:binary_res_confusion} for the spectrogram feature and the best performing wav2vec feature (wav2vec-1). It can be seen that there are less confusions between healthy and dysarthric speech for the wav2vec-1 feature compared to the spectrogram feature.

\begin{table}[]
\centering
\caption{\label{tab:binary_res} Dysarthria detection results for the three baseline features along with the best three wav2vec features. Here ACC refers to accuracy, SE refers to sensitivity, SP refers to specificity, and F1 refers to F1-score.}
\vspace{0.1cm}
\begin{tabular}{|l|l|l|l|l|}
\hline
\multicolumn{1}{|c|}{{\color[HTML]{000000} Feature}} & \multicolumn{1}{c|}{{\color[HTML]{000000} ACC [\%]}} & \multicolumn{1}{c|}{{\color[HTML]{000000} SE}} & \multicolumn{1}{c|}{{\color[HTML]{000000} SP}} & \multicolumn{1}{c|}{{\color[HTML]{000000} F1}} \\ \hline \hline
\multicolumn{5}{|c|}{{\color[HTML]{000000} \textbf{Baseline features}}} \\ \hline \hline

{\color[HTML]{000000} Spectrogram}                   & {\color[HTML]{000000} 92.72}                    & {\color[HTML]{000000} 0.92}                    & {\color[HTML]{000000} 0.94}                    & {\color[HTML]{000000} 0.93}                    \\ \hline
{\color[HTML]{000000} Mel-spectrogram}               & {\color[HTML]{000000} 91.48}                    & {\color[HTML]{000000} 0.91}                    & {\color[HTML]{000000} 0.92}                    & {\color[HTML]{000000} 0.92}                    \\ \hline
{\color[HTML]{000000} MFCCs}                         & {\color[HTML]{000000} 89.23}                    & {\color[HTML]{000000} 0.86}                    & {\color[HTML]{000000} 0.93}                    & {\color[HTML]{000000} 0.90}                    \\ \hline \hline
\multicolumn{5}{|c|}{{\color[HTML]{000000} \textbf{Wav2vec features}}} \\ \hline \hline
{\color[HTML]{000000} wav2vec-1}                  & {\color[HTML]{000000} {\bf 93.95}}                    & {\color[HTML]{000000} {\bf 0.93}}                    & {\color[HTML]{000000} {\bf 0.95}}                    & {\color[HTML]{000000} {\bf 0.94}}                    \\ \hline
{\color[HTML]{000000} wav2vec-2}                  & {\color[HTML]{000000} 91.96}                    & {\color[HTML]{000000} 0.90}                    & {\color[HTML]{000000} 0.94}                    & {\color[HTML]{000000} 0.92}                    \\ \hline
{\color[HTML]{000000} wav2vec-3}                  & {\color[HTML]{000000} 91.97}                    & {\color[HTML]{000000} 0.90}                    & {\color[HTML]{000000} 0.95}                    & {\color[HTML]{000000} 0.92}                    \\ \hline
\end{tabular}
\vspace{-0.1cm}
\end{table}

\begin{table}[]
\centering
\caption{\label{tab:binary_res_confusion} Confusion matrices of dysarthria detection for the spectrogram feature (the best performing baseline) along with wav2vec-1 (the best performing wav2vec feature).}
\vspace{0.1cm}
\resizebox{\columnwidth}{!}{%
\begin{tabular}{|ccc|lcc|}
\hline
\multicolumn{3}{|c|}{{\color[HTML]{000000} Spectrogram}}                                                              & \multicolumn{3}{c|}{wav2vec-1}                                                                \\ \hline
\multicolumn{1}{|c|}{}                                  & \multicolumn{1}{c|}{Healthy} & Dysarthric                   & \multicolumn{1}{c|}{}           & \multicolumn{1}{c|}{Healthy}                      & Dysarthric \\ \hline
\multicolumn{1}{|l|}{{\color[HTML]{000000} Healthy}}    & \multicolumn{1}{c|}{93.94}   & {\color[HTML]{000000} 6.06}  & \multicolumn{1}{l|}{Healthy}    & \multicolumn{1}{c|}{{\color[HTML]{000000} 94.85}} & 5.15       \\ \hline
\multicolumn{1}{|c|}{{\color[HTML]{000000} Dysarthric}} & \multicolumn{1}{c|}{8.34}    & {\color[HTML]{000000} 91.66} & \multicolumn{1}{l|}{Dysarthric} & \multicolumn{1}{c|}{{\color[HTML]{000000} 6.82}}  & 93.18      \\ \hline
\end{tabular}}
\vspace{-0.2cm}
\end{table}

\subsection{Severity level classification of dysarthric speech}
\label{sec:res_multi-class}
The performance in term of accuracy are shown in Figure~\ref{fig:fig3} for the three baseline features and for all the 13 wav2vec features for the severity level classification experiments. Table~\ref{tab:multi_class_res} shows the overall classification accuracy together with the class-wise accuracies for the baseline features and for the two best wav2vec features. 

From Figure~\ref{fig:fig3}, it can be clearly seen that almost all the wav2vec features (except for wav2vec-1) outperformed the three baseline features in terms of the mean accuracy. Interestingly, unlike in the detection problem, the best-performing wav2vec features were the ones obtained from the final layers. In fact, there is a rising trend in the accuracy when moving from the first layer towards the final layer. This result was expected because the severity of dysarthria is associated with the intelligibility of speech (e.g., phoneme identity) and the wav2vec pre-trained model is designed for the ASR tasks, therefore the final layers can effectively learn information related to the linguistic contents of speech. 

Among the baseline features, it can be observed that the MFCCs and spectrogram performed better than the mel-spectrogram (chance level is 25\%). Compared to the best baseline (MFCCs), wav2vec-12 and wav2vec-13 gave an absolute improvement of 8.88\% and 10.62\%, respectively. The results also show that for the wav2vec features, the class-wise accuracies are relatively less biased towards the two extreme ends of the severity scale ("very low" and "high" levels of dysarthria severity) compared to the baseline features. 

\begin{table}
\centering
\caption{\label{tab:multi_class_res} Dysarthria severity level classification accuracies and class-wise accuracies for the three baseline features along with the two best wav2vec features. Here ACC refers to accuracy and C refers to class.}
\vspace{0.1cm}
\resizebox{\columnwidth}{!}{
\begin{tabular}{|l|l|l|l|l|l|}
\hline
\multicolumn{1}{|c|}{{\color[HTML]{000000} Feature}} & \multicolumn{1}{c|}{{\color[HTML]{000000} ACC [\%]}} & \multicolumn{1}{c|}{{\color[HTML]{000000} $C_{very-low}$}} & \multicolumn{1}{c|}{{\color[HTML]{000000} $C_{low}$}} & \multicolumn{1}{c|}{{\color[HTML]{000000} $C_{medium}$}} & \multicolumn{1}{c|}{{\color[HTML]{000000} $C_{high}$}} \\ \hline \hline
\multicolumn{6}{|c|}{{\color[HTML]{000000} \textbf{Baseline features}}} \\ \hline \hline

{\color[HTML]{000000} Spectrogram}                   & {\color[HTML]{000000} 33.26}                    & {\color[HTML]{000000} 44.54}                    & {\color[HTML]{000000} 16.65}                    & {\color[HTML]{000000} 14.95}                    & {\color[HTML]{000000} 56.87} \\ \hline
{\color[HTML]{000000} Mel-spectrogram}               & {\color[HTML]{000000} 26.21}                    & {\color[HTML]{000000} 32.10}                    & {\color[HTML]{000000} 13.71}                    & {\color[HTML]{000000} 16.09}                     & {\color[HTML]{000000} 42.94} \\ \hline
{\color[HTML]{000000} MFCCs}                         & {\color[HTML]{000000} 33.94}                    & {\color[HTML]{000000} 50.63}                    & {\color[HTML]{000000} 7.32}                    & {\color[HTML]{000000} 21.02}                     & {\color[HTML]{000000} 56.76} \\ \hline \hline
\multicolumn{6}{|c|}{{\color[HTML]{000000} \textbf{Wav2vec features}}} \\ \hline \hline
{\color[HTML]{000000} wav2vec-12}                  & {\color[HTML]{000000} {\bf 42.82}}                    & {\color[HTML]{000000} 63.10}                    & {\color[HTML]{000000} 22.77}                    & {\color[HTML]{000000} 16.51}                     & {\color[HTML]{000000} 68.91} \\ \hline
{\color[HTML]{000000} wav2vec-13}                  & {\color[HTML]{000000} {\bf 44.56}}                    & {\color[HTML]{000000} 56.09}                    & {\color[HTML]{000000} 23.21}                    & {\color[HTML]{000000} 18.77}                     & {\color[HTML]{000000} 80.16} \\ \hline
\end{tabular}}
\vspace{-0.2cm}
\end{table}

\section{Summary and Conclusions}
In this study, we explored the state-of-the-art pre-trained wav2vec model to extract features in the context of dysarthric speech detection and in the context of severity level classification of dysarthria. 
A comparison of the wav2vec features with the popularly used spectral and cepstral-based baseline features (spectrogram, mel-spectrogram, and MFCCs) was carried out both in the detection problem and in the severity level classification problem. 

The results of the dysarthric speech detection experiments indicated that the features extracted from the first layer (wav2vec-1) outperformed the baseline and other wav2vec features by showing an absolute improvement of 1.23\% in accuracy compared to the best performing baseline feature (spectrogram). This indicates that the starting layers of the wav2vec model has learned generic speech features that can be effectively used for the detection of dysarthric speech. The results of the severity level classification experiments showed that the features extracted from the final layers (wav2vec-12 and wav2vec-13) performed better than the baseline and the other wav2vec features. Compared to MFCCs (the best baseline feature), an absolute improvement of 8.88\% and 10.62\% was given by wav2vec-12 and wave2vec-13, respectively. We argue that the improved performance that was obtained using the features from the final layers is due to the wav2vec model's capability to extract features that contain information related to the linguistic contents associated to speech intelligibility (which is directly related to the severity level of dysarthria).

Taken together, the experimental findings of the study indicate that the classification systems seem to be generalizable to unseen speakers using the features from the starting layers in the detection task, and the features from the final layers in the multi-class classification task. However, further research is required in order to study the generalizability of the wav2vec features for other disorders and to study also the performance of these features in cross-database scenarios.




\end{document}